\documentclass[aps,prl,showpacs,floatfix,preprint,superscriptaddress]{revtex4-1}

\usepackage{graphicx,color}
\usepackage{amssymb,amsthm,amsmath}
\begin{document}

\title{Electric field induced permanent superconductivity in layered metal nitride chloride $M$NCl($M:$ Hf and Zr)}


\author{Shuai Zhang}
\email{szhang@iphy.ac.cn}
\affiliation{Institute of Physics and Beijing National Laboratory for Condensed Matter Physics, Chinese Academy of Sciences, Beijing 100190, China} 

\author{Moran Gao}
\affiliation{Institute of Physics and Beijing National Laboratory for Condensed Matter Physics, Chinese Academy of Sciences, Beijing 100190, China}
\affiliation{School of Physical Sciences, University of Chinese Academy of Sciences, Beijing 100049, China}
\author{Huanyan Fu}
\affiliation{Institute of Physics and Beijing National Laboratory for Condensed Matter Physics, Chinese Academy of Sciences, Beijing 100190, China}
\affiliation{School of Physics and Electronics, Shandong Normal University, Jinan 250014, China}
\author{Xinmin Wang}
\affiliation{Institute of Physics and Beijing National Laboratory for Condensed Matter Physics, Chinese Academy of Sciences, Beijing 100190, China}
\affiliation{School of Physical Sciences, University of Chinese Academy of Sciences, Beijing 100049, China}
\author{Zhian Ren}
\affiliation{Institute of Physics and Beijing National Laboratory for Condensed Matter Physics, Chinese Academy of Sciences, Beijing 100190, China} 
\affiliation{School of Physical Sciences, University of Chinese Academy of Sciences, Beijing 100049, China}
\affiliation{Collaborative Innovation Center of Quantum Matter, Beijing 100190, China}
\author{Genfu Chen}
\affiliation{Institute of Physics and Beijing National Laboratory for Condensed Matter Physics, Chinese Academy of Sciences, Beijing 100190, China} 
\affiliation{School of Physical Sciences, University of Chinese Academy of Sciences, Beijing 100049, China}
\affiliation{Collaborative Innovation Center of Quantum Matter, Beijing 100190, China}

\begin{abstract}
The device of electric double-layer transistor (EDLT) with ionic liquid has been employed as an effective way to dope carriers over a wide range, which can induce metallic state \cite{YuanAFM2009,LengNPJ2017}, magnetic reconstruction\cite{YuanNP2013,ZhangNC2014} and superconducting transition\cite{UenoNM2008,YeNM2009,YeScience2012}. However, the induced electronic state can hardly survive in the materials after releasing the gate voltage, strongly restricting the experimental study for discovery of physics. Here, we show that a permanent superconductivity with transition temperature $T_c$ of 24 and 15 K is realized in single crystals and polycrystalline samples of HfNCl and ZrNCl upon applying proper gate voltage $V_{\rm G}$, respectively. Reversible change between insulating and superconducting state can be obtained through applying positive and negative $V_{\rm G}$ at low temperature such as 220 K, whereas $V_{\rm G}$ applied at high temperatures ($\geq$ 250 K) could induce partial deintercalation of Cl ions and result in irreversible superconducting transition. The present results clarify a connection between traditional chemical synthesis and the electrochemical mechanism of the EDLT induced superconductivity. Such a technique shows a great potential to systematically tune the bulk electronic state in the similar two-dimensional (2D) systems.
\end{abstract}

\maketitle

\section{Introduction}
Due to the advantage of easy fabrication process, EDLT devices with liquid dielectric has been widely used to tune the carrier density of 2D materials over a wide doping range. The carrier doped phase always shows dramatic change as compared to parent phase. As one of the most complex physics phenomena, EDLT induced superconducting transition from insulating parent phase has already become to an important topic in condensed matter research. Previous findings in SrTiO$_3$\cite{UenoNM2008}, ZrNCl\cite{YeNM2009} and MoS$_2$\cite{YeScience2012} provide sufficient information and ideas on device design and key parameters such as liquid dielectric and gate voltage $V_{\rm G}$. However, several fundamental requirements have to be satisfied to achieve the functional liquid/solid interface. First, mechanical micro-cleavage technique is used to prepare untrathin film with high-quality surface because the carrier depleting and accumulating can occur only on surface. Second, electrodes with micro size are patterned for electrical transport measurements. Third, as the gate dielectric, proper types of ionic liquids are chosen for applying $V_{\rm G}$ without destroying the sensitive crystal surface. Most importantly, the induced superconducting state remains no longer in the system after releasing the gate voltage. All these severe requirements confine the applying of EDLT method to other systems. 

In recent studies, EDLT device with liquid dielectric has turned into a more powerful technique, which can induce not only the continuous doping of two types of carriers but also the structure transformation. The detailed tuning through the charge density wave (CDW) and superconductivity phase in 1T-TiSe$_2$ classified the universality of quantum critical point (QCP)\cite{LiNature2016}. A tri-state phase transformation was realized in SrCoO$_{2.5}$ by using an electric field to control the insertion and extraction of oxygen and hydrogen ions electrolyzed from H$_2$O\cite{LuNature2017}. The local carrier doping is generally caused by the movement of carriers from the inside of crystal to surface. Upon applying gate voltage, external ions in the dielectric liquid as well as the carriers (electrons or holes) in the crystal can be completely controlled to tune the electronic states. 

Upon electron doping by intercalation of alkali, alkaline-earth and rare-earth metals, metal nitride halides $M$N$X$ ($M$: Ti, Zr, Hf; $X$: Cl, Br, I) show bulk superconductivity with relatively high $T_c$ values, 18 K in $\alpha$-TiNCl\cite{ZhangPRB2012}, 26 and 15.5 K for $\beta$-HfNCl\cite{Yamanaka1998,ZhangSUST20131,ZhangSUST20132} and ZrNCl\cite{TakanoPRL2008}, respectively. In this study, we carried out detailed experimental studies on layered metal nitride chloride $M$NCl ($M$: Hf and Zr) using EDLT technique (Supplementary Fig. S1). Permanent superconductivity with $T_c=$ 24 and 15 K were realized on HfNCl and ZrNCl, respectively, strongly implying that electrical-double-layer state could be an intermediate stage. It appears that the EDLT induced partial deintercalation of Cl is the intrinsic electrochemical mechanism of the permanent superconductivity observed in layered metal nitride chloride.

\section{Experimental details} 
Pristine HfNCl and ZrNCl single crystals were grown using a well-established chemical transport method\cite{Yamanaka1998}. Typical crystal size is 300$\times$200$\times$20 $\mu m$. Layered crystals were fixed on a SiO$_2$ surface grown on Si substrate. Normal silver paste was used to set electrodes for electric transport measurements. Lots of tiny single crystals were pressed into pellets with thickness of 0.1$\sim$0.2 mm, which were used as polycrystalline sample. Three types of ionic liquid were chosen as the gate dielectric; Diethylmethyl (2-methoxyethyl) ammonium bis (trifluoromethylsulfonyl) imide (DEME-TFSI), (EMIM-TFSI ), and 1-Ethyl-3-methylimidazolium tetrafluoroborate (EMIM-BF$_4$). Gate voltage was applied up to $\pm 5.5$ V at several temperatures between 220 and 300 K. The applied gate voltage $V_{\rm G}$ was kept as cooling down the system and released after the appearance of $\pm$ 0.1 nA of leakage current. The electrical resistance $R(T)$s discussed in the paper were measured in a warming up process.

 The magnetic susceptibility on the same single crystals was measured after completing the electric transport measurements. After reapplying proper gate voltages at corresponding temperatures (5.5 V at 220 K and 3.5 V at 250 K), the system was warmed up to room temperature and the same single crystals were transferred to the chamber for the magnetization measurement using a SQUID magnetometer (Quantum Design SQUID VSM). Because of the unmeasurable mass of single crystals ($\leq 0.1$ mg), the slight error was caused due to the weak magnetization signal, which is close to the measurement accuracy. However, the superconducting transitions can still be clearly determined as shown in Fig. \ref{f2}b and Fig. \ref{f5}c.

\section{Results and discussion}
\begin{figure}
\centering
\includegraphics[width=1\linewidth]{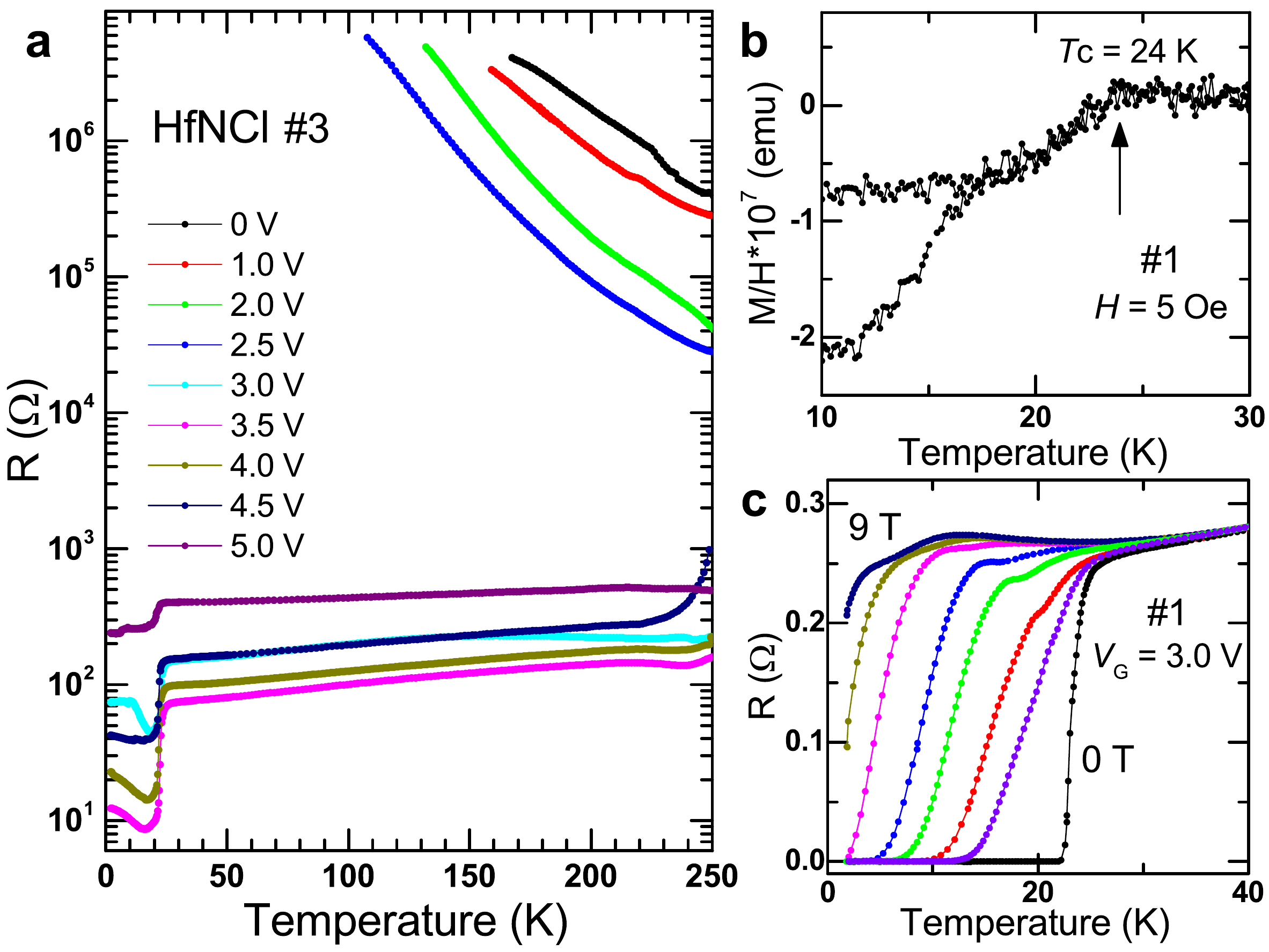}%
\caption{\label{f2} EDLT induced permanent superconductivity with $T_c=$ 24 K in HfNCl single crystals. a, Temperature dependence of resistance $R(T)$ at different gate voltage $V_{\rm G}$ from 0 to 5 V. b, Temperature dependence of magnetic susceptibility $\chi$ of the same single crystal ($\sharp 1$) used in $R(T)$ measurements. c, Temperature dependence of $R(T)$ under different magnetic fields ($\mu_{0}H=$ 0, 0.5, 1, 2, 3, 5, 7, 9 T) at $V_{\rm G}=$ +3.0 V.}
\end{figure}

Figure \ref{f2}a shows the temperature dependence of the resistance $R(T)$ at different $V_{\rm G}$. The $R(T)$ curves line up in two groups; typical insulating $R(T)$ for $V_{\rm G}\leq 2.5$ V and metallic $R(T)$ for $V_{\rm G}\geq 3$ V. For the insulating $R(T)$ groups, the absolute value of resistance decreases monotonically with increasing $V_{\rm G}$, implying the increase of carrier density in the system. An apparent insulator-metal transition with the decrease of resistance in four orders of magnitude occurs between $V_{\rm G}=$ 2.5 and 3 V. The steep drop around 25 K observed in $R(T)$ with $V_{\rm G}\geq 3$ V is the sign of superconducting transition, which was confirmed in magnetic susceptibility $\chi(T)$ (Fig. \ref{f2}b) and $R(T)$ under magnetic fields (Fig. \ref{f2}c). The transition temperature $T_c$ is determined as 24 K (Supplementary Fig. S2), and the zero resistance at 22.2 K is higher than all the values reported for intercalated superconductors $A_x$HfNCl with/without cointercalation of molecules\cite{Yamanaka1998,TakanoPRL2008,ZhangSUST20131}. The confirmation of $T_c$ in $\chi(T)$ proves directly the permanent characteristic of the superconducting state without applying gate voltage, which is strongly contrast to the general EDLT induced superconductivity. 

\begin{figure}
\centering
\includegraphics[width=15cm]{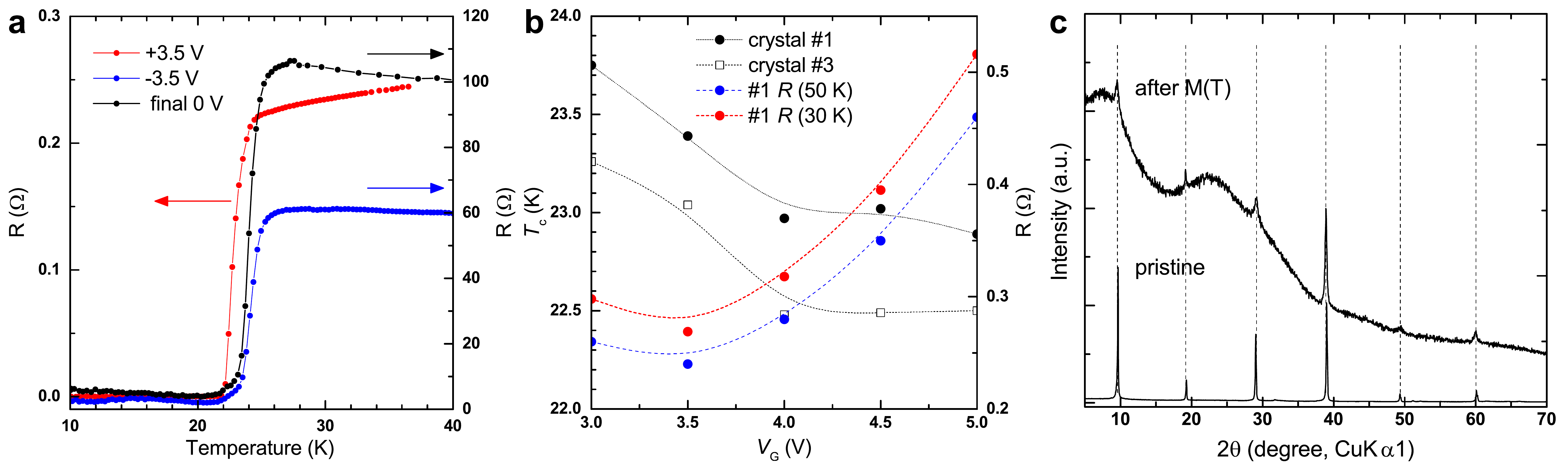}%
\caption{\label{f3} Permanent superconductivity in HfNCl after releasing gate voltage. a, Temperature dependence of electrical resistance $R(T)$ with applying positive, negative and zero $V_{\rm G}$. b, Gate voltage dependence of the induced $T_c$ (left scale) and absolute resistance $R(30 \rm K)$ and $R(50 \rm K)$ (right scale). c, XRD pattern of the same layered HfNCl crystal before and after all measurements with EDLT function.}
\end{figure}

To dynamically probe the EDLT induced superconducting state, gate voltage of -3.5 and 0 V was applied after completing the transport measurements under $V_{\rm G}=$ 5 V. As compared to the $R_{V_{\rm G}=3.5 V}(T)$ in Fig. \ref{f3}a, the onset of $T_c$ increases up to near 27.3 K although the absolute resistance is increased by two orders of magnitude. Note that the superconducting transition remains in the material after all the transport measurements. Figure \ref{f3}b shows the $V_{\rm G}$ dependence of transition temperature $T_c$ observed on single crystals $\#1$ and $\#3$. As increasing $V_{\rm G}$ from 3 to 5 V, near 1 K weakening of $T_c$ was confirmed on all selected crystals. The $V_{\rm G}$ dependence of $T_c$ is consistent with the $T_c$ depending on doping level observed on bulk intercalated superconductors such as Li$_xM_y$HfNCl\cite{TakanoPRL2008} and $AE_xM_y$HfNCl\cite{ZhangSUST20131}. In the meanwhile, the resistance $R(30 \rm K)$ and $R(50 \rm K)$ starts to increase as increasing $V_{\rm G}$ above 3.5 V, which suggests that optimal gate voltage for the bulk conductivity occurs around this area. As shown in Fig. \ref{f3}c, the [00l] reflection peak in XRD pattern shows no apparent shift compared to the pristine single crystals, excluding the possibility of intercalation of big-size ions from liquid dielectric. 

\begin{figure}
\centering
\includegraphics[width=1\linewidth]{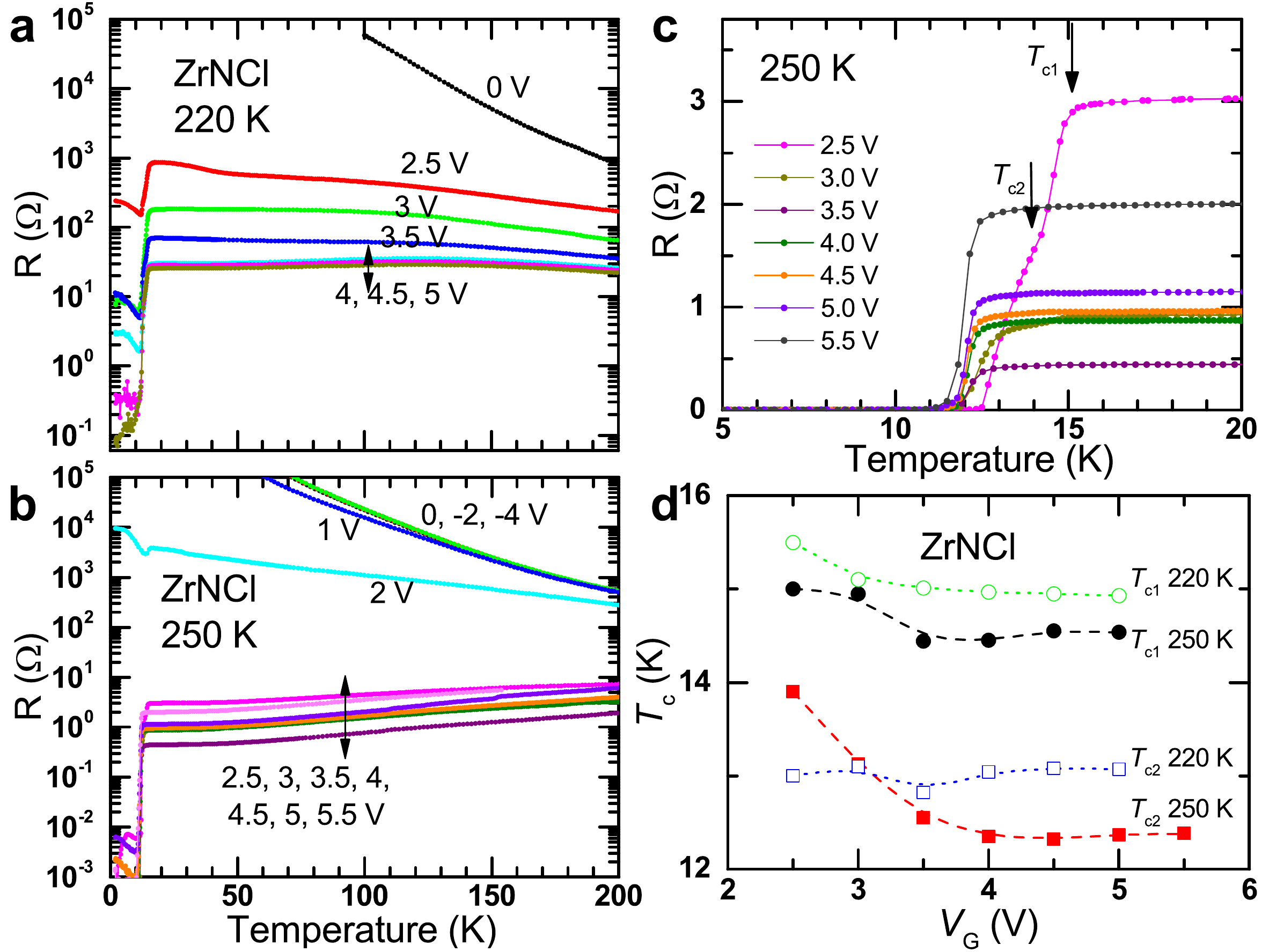}%
\caption{\label{f4} Intrinsic difference of the the superconductivity induced by applying gate voltage at different temperatures. a,b, Systematic temperature dependence of resistance with applying gate voltage at 220 and 250 K, respectively. c, The $T_c$ is clearly identified in low-temperature $R(T)$. d, $V_{\rm G}$ dependence of induced $T_{c1}$ and $T_{c2}$.}
\end{figure}

To clarify the difference of intrinsic mechanism between the present findings on bulk single crystals and previous studies on atomically flat film of ZrNCl\cite{YeNM2009,SaitoNC2018}, we verify the EDLT function on ZrNCl with applying gate voltage at 220 (LT) and 250 K (HT). Figure \ref{f4}a and \ref{f4}b show systematic $R_{\rm HT}(T)$ and $R_{\rm LT}(T)$ behaviors at different $V_{\rm G}$, respectively. At 220 K, EDLT induced insulating-superconducting transition around 15 K is in great agreement with previous results observed on atomically flat film; the absolute resistance decreases monotonically with increasing $V_{\rm G}$ and negative gate voltage can accelerate the restoring of $R(T)$ to an insulating-like behavior. Without gate voltage, the resistance at 220 K continuesly increases as extending the relaxation time (Supplementary Fig. S3). For 250 K, on the other hand, $R_{\rm HT}(T)$ at $V_{\rm G}=-2$ and $-4$ V shows no change compared to $V_{\rm G}=0$ V. As increasing positive $V_{\rm G}$, absolute resistance decreases monotonically between 0 and 3.5 V, and increases again from 3.5 to 5.5 V. The overall trendency of $R_{\rm HT}(T)$ reproduces fully the observations on HfNCl discussed previously (Fig. \ref{f3}b), strongly implying that the temperature at which gate voltages were applied plays a dominant role in EDLT function. The transition temperature $T_c$ is clearly determined in low-temperature $R(T)$ as shown in Fig. \ref{f4}c. At $V_{\rm G}=$ 2.5 V, two clear transitions were confirmed at $T_{c1}$ and $T_{c2}$, which are similar to the previous observations on ZrNCl\cite{YeNM2009}. We consider that such a two-step transition is caused by the nonhomogeneity of the transport channel. The $T_{c1}$ and $T_{c2}$ values determined in $R(T)$ with applying gate voltage at both 220 and 250 K are summarized in Fig. \ref{f4}(d). All values of $T_c$ display a rather weak $V_{\rm G}$ dependence. $T_c$ decreases monotonically with increasing $V_{\rm G}$, showing an agreement with the carrier doping level dependence of $T_c$ observed on bulk intercalated superconductors\cite{TaguchiPRL2006}.

\begin{figure}
\centering
\includegraphics[width=1\linewidth]{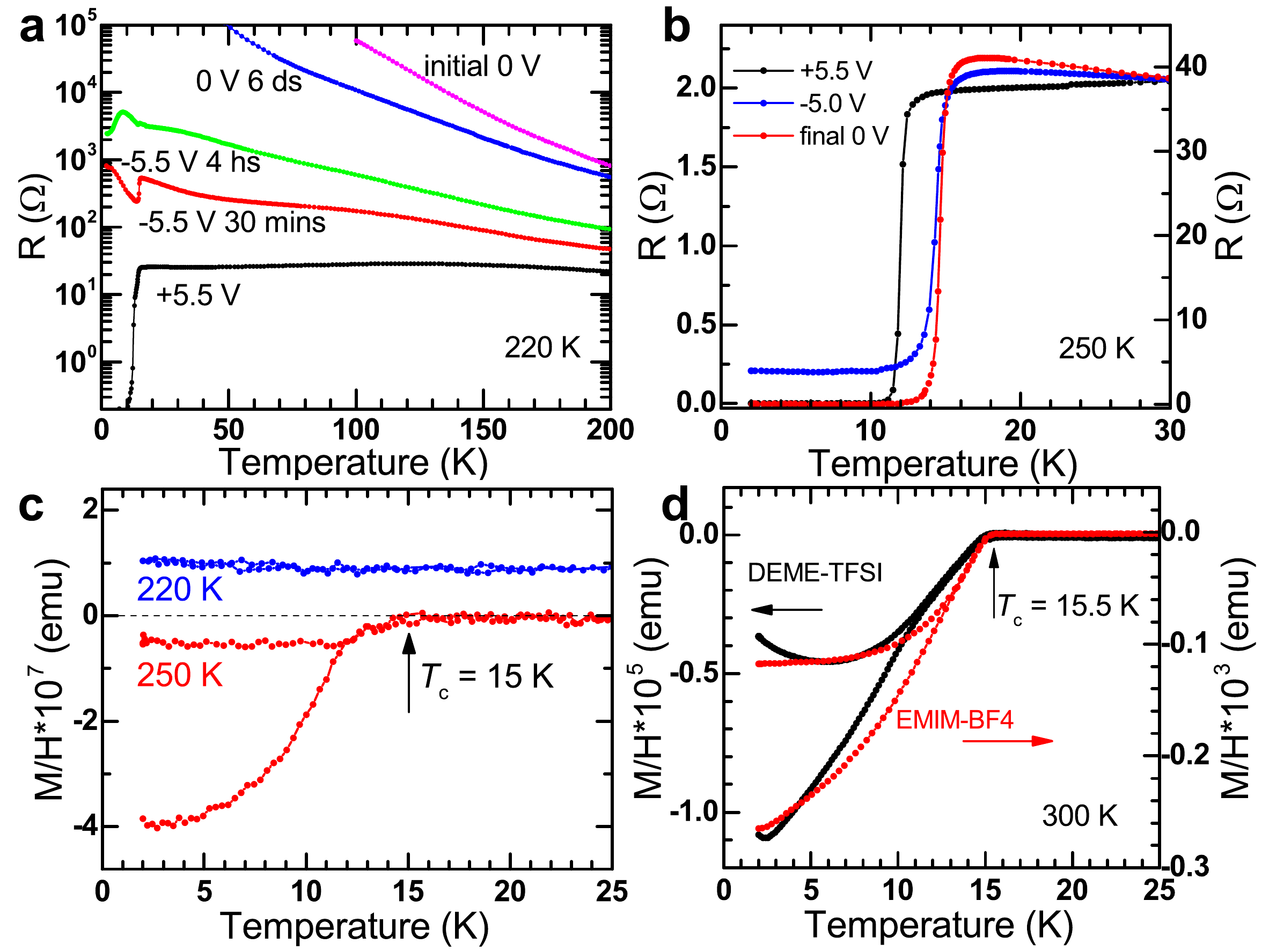}%
\caption{\label{f5} Reversible and irreversible $R(T)$ with applying gate voltage at different temperatures. a,b, At 220 and 250 K, respectively, temperature dependence of resistance $R$ shows reversible and irreversible behaviors as applying negative gate voltages. c, Temperature dependence of magnetic susceptibility $M/H(T)$ of the same single crystal used in $R(T)$ measurements. c, The EDLT function is successfully extended to big-size polycrystalline samples using different ionic liquid. $T_c$ of 15.5 K is confirmed in magnetic susceptibility $M/H(T)$.}
\end{figure}

Negative voltages were applied to probe the difference between applying gate voltage at 220 and 250 K (Fig. \ref{f5}a and b). Upon applying $V_{\rm G}=$ -5 V at 220 K for several hours after measurements at $V_{\rm G}=$ +5.5 V , $R_{\rm LT}(T)$ could be restored to thousands of ohm at low temperatures. Several-day relaxation without gate voltage at room temperature could restore $R(T)$ to the insulating state. Although the continuous recovery from superconducting to insulating state is quite time-consuming, the induced electronic state disappears as releasing the gate voltage, which is similar to the general EDLT induced behavior. In contrast, the $T_c$ induced by applying $V_{\rm G}$ at 250 K increases up to $\sim$15.3 K with applying $-5$ V or long-time relaxation without gate voltage at the same temperature. In the meanwhile, the absolute value of $R_{\rm HT}(T)$ shows an increase of more than ten times as well as the disappearance of the two-step transition. We conclude that the applying of negative gate voltage contributes to the relaxation of electronic state, which can lead to bulk feature with more homogeneity. Similar behavior was also observed on HfNCl (Fig  \ref{f3}a).

 As a double check, temperature dependence of magnetic susceptibility $\chi(T)$ was measured on the same single crystals (see methods). $\chi(T)$ shows a clear superconducting transition around 15 K for $V_{\rm G}$ applied 250 K whereas no transition was found down to 1.8 K for $V_{\rm G}$ applied at 220 K, implying the permanent superconductivity in the former case (Fig \ref{f5}c). The upper critical field $H_{\rm c2}$ was estimated using the temperature dependence of resistance $R(T)$ in different fields (Supplementary Fig. S4). No apparent difference was confirmed between the $H_{\rm c2}(T)$ for superconducting states induced at 220 and 250 K, showing the similar response to external magnetic fields. Furthermore, taking the advantage of the layered characteristic in the present system, we have successfully extended the EDLT function to the polycrystalline samples using different type of ionic liquids (DEME-TFSI, EMIM-TFSI and EMIM-BF4). As shown in Fig. \ref{f5}d, the induced superconducting transitions are consistent with the bulk superconductivity observed on intercalated superconductors.

\begin{figure}
\centering
\includegraphics[width=15cm]{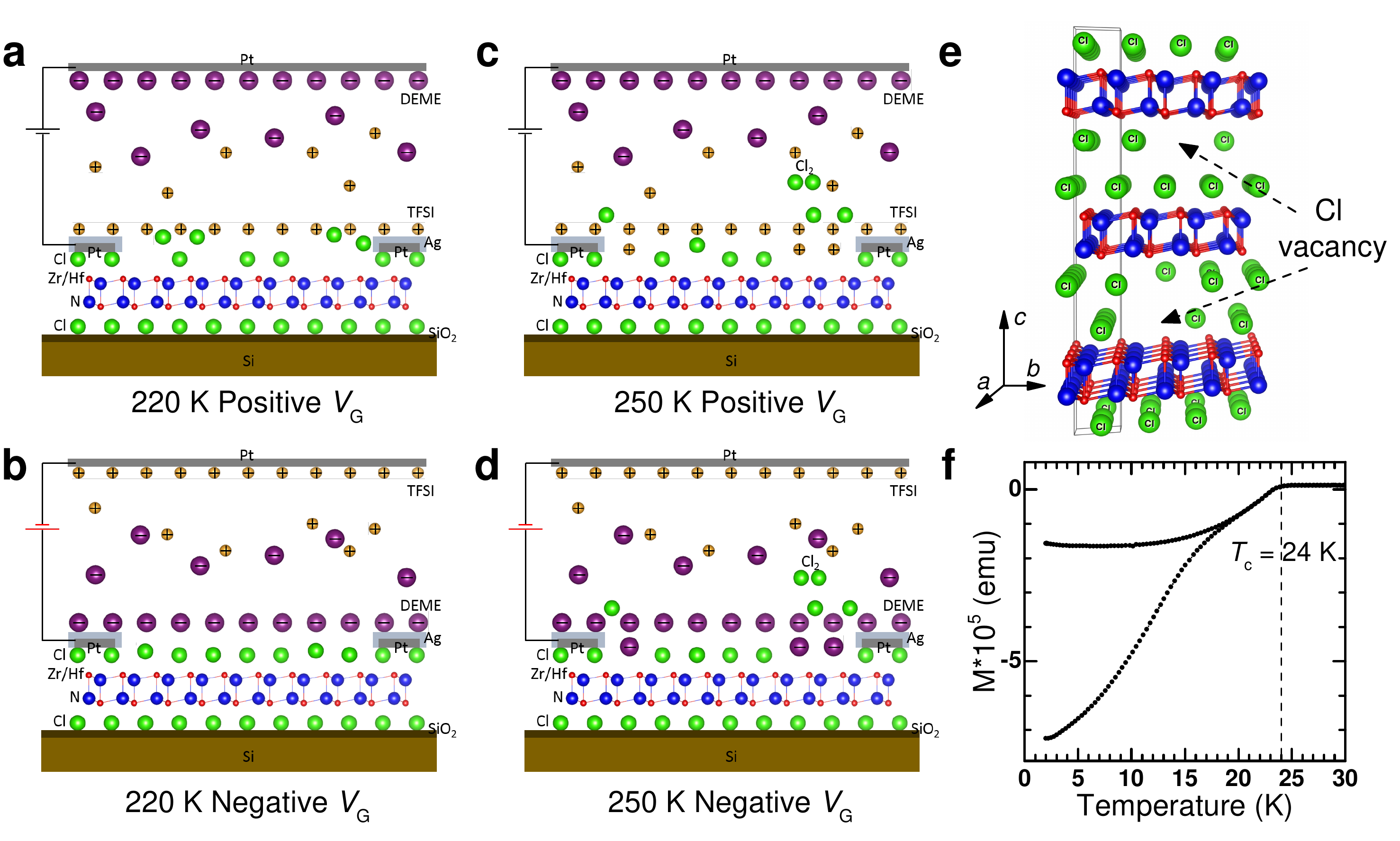}%
\caption{\label{f7} Concept of EDLT induced phase transformation. a,b, Reversible  phase transformation obtained by applying positive and negative gate voltage at 220 K. c,d, Irreversible phase transformation occurs as applying gate voltage at higher temperatures such as 250 K. e, Structure model of the EDLT induced phase. f, A typical SC state observed on polycrystalline samples of HfNCl. Upon applying EDLT function on polycrystalline samples, the diamagnetism corresponding to SC transition was observed in temperature dependence of magnetic susceptibility $M/H(T)$.}
\end{figure}

Here, we propose a reasonable scenario interpreting the difference between applying gate voltage at 220 and 250 K. Upon applying gate voltage at 220 K (Fig. \ref{f7}a), short-range movement of partial Cl$^{1-}$ ions could be induced by the positive $V_{\rm G}$, leading to a local electron doping in the system. The applying of negative $V_{\rm G}$ can almost push back the Cl ions to the initial positions (Fig. \ref{f7}b). This is consistent to the induced superconducting transition and restored insulating state (Fig. \ref{f5}a). On the other hand, as applying gate voltage at 250 K, partial Cl$^{1-}$ ions could get enough energy to escape from the crystal surface (Fig. \ref{f7}c). The boiling temperature of Cl$_2$ is near 240 K. The possible formation of Cl$_2$ can provide extra electrons, which results in a permanent electron doping to the system. Correspondingly, the negative $V_{\rm G}$ can only contribute to the improvement of the homogeneity more than pushing back the Cl ions to the system (Fig. \ref{f7}d). Thus, the permanent superconductivity observed in the present study is caused by the partial deintercalation of Cl, suggesting that the EDLT function induced liquid/solid interface is an intermediate stage. Similar process was ever reported on VO$_2$ film, in which suppression of metal-insulator transition was caused by electric field-induced oxygen vacancy formation\cite{JeongScience2013}.

As discussed before, for both HfNCl and ZrNCl, negative gate voltages could result in the increase of $T_c$ and absolute resistance $R(T)$ (Fig. \ref{f3}a and Fig. \ref{f5}b), which can also be interpreted by the model. The poor homogeneity of the transport channel is caused by the random deintercalation of Cl. Upon applying negative gate voltage or long-time relaxation at high temperature without gate voltage, the local high density of Cl vacancies turn into relatively uniform low density of vacancies, which is equal to the change from local high doping level to uniform low carrier density. Correspondingly, the $T_c$ increases as decreasing carrier density, which is consistent with the observations on bulk superconductors\cite{TakanoPRL2008}. In the meanwhile, the local strong scattering turns into a relatively uniform weak scattering, which is responsible for the increase of absolute $R(T)$. We consider that such a process is a dynamic relaxation especially under negative gate voltages.

Upon optimization of all variable parameters in the EDLT function, we have obtained a clear and permanent superconducting states in HfNCl as shown in Fig. \ref{f7}f. As already mentioned before, the EDLT induced superconducting state occurs mainly on crystal surface, and has poor homogeneity. Thus, parital deintercalation of Cl occurs randomly on the crystal surfaces, which is also evidenced by the absent color change of the crystals after processing EDLT function. The deintercalation with a relative uniform distribution could result in a primary phase of HfNCl$_{1-x}$ (Fig. \ref{f7}e), which shows the superconducting transition at 24 K as confirmed in the temperature dependence of magnetization (Fig. \ref{f7}f).  Similar behavior was already reported on bulk deintercalated samples\cite{ZhuCM2003}. Such a experimental fact supports strongly the scenario of EDLT induced Cl deintercalation at high processing temperatures. 

In summary, we unfold a connection between traditional chemical synthesis and EDLT induced liquid/solid interface on layered $M$NCl ($M:$ Hf and Zr) crystals. Compared to the strongly bonded honeycomb-like double $M$N layer, the interaction between chlorine and $M$N layers is rather weak. Upon applying positive and negative gate voltage at low temperature such as 220 K, system shows a reversible change between insulating and superconducting state, whereas the EDLT function applied at high temperatures ($\geq$ 250 K) results in the irreversible superconductivity with $T_c=$ 24 and 15 K on HfNCl and ZrNCl, respectively. This findings imply that the EDLT function induced partial deintercalation of Cl ions is the intrinsic nature for the permanent superconductivity. The discovery of such an electrochemical mechanism of permanent superconductivity will caused wide reconsideration of similar phenomena reported before, and shed light on tuning bulk electronic state in similar 2D systems using devices with EDLT function.




\section*{Acknowledgments}
 This work was supported by the National Natural Science Foundation of China (Grant No. 11704403), the National Key Research Program of China (Grant No. 2016YFA0401000 and 2016YFA0300604) and the Strategic Priority Research Program (B) of Chinese Academy of Sciences (Grant No. XDB07020100).

S.Z conceived the project and designed the experiments. X.W. and H.F. fabricated the polycrystal samples and single crystals of Zr/HfNCl. S.Z. designed and fabricated the EDLT devices. S.Z., M.G. and H.F. carried out all of the measurements. S.Z. analysed the data and wrote the paper, and all authors commented on it. G.C. provided fruitful discussions.   

Correspondence and requests for materials should be addressed to S.Z.(email: szhang@iphy.ac.cn) and G.C. (email: gfchen@iphy.ac.cn).

\end{document}